\documentstyle[aps,prd,preprint]{revtex}
\def\be{\begin{equation}}
\def\ee{\end{equation}}
\def\beq{\begin{eqnarray}}
\def\eeq{\end{eqnarray}}

\def\R{{\cal R}}

\def\P{\partial}
\def\nn{\nonumber}

\begin{document}
\begin{titlepage}
\begin{flushright}  
\end{flushright}
\begin{center}
  {\bf \LARGE STEALTH BRANES} 
\vskip 2em
{\large Rub\'en Cordero\footnote{${}^{}$ On leave from Departamento de F\'{\i}sica, 
Escuela Superior de F\'{\i}sica y Matem\'aticas 
del IPN 
Edificio 9, 07738, M\'exico D.F., MEXICO}
 and 
Alexander Vilenkin${}^{}$ \\[1em]}
\em{
${}^{}$ Institute of Cosmology, Department of Physics and Astronomy,\\
Tufts University, Medford, Massachusetts, 02155, USA}\\[1em]
\end{center}

\begin{abstract}

We discuss the brane world model of Dvali, Gabadadze and Porrati in
which branes evolve in an infinite bulk and the brane curvature term
is added to the action.  If $Z_2$ symmetry between the two sides of
the brane is not imposed, we show that the model admits the existence
of ``stealth branes'' which follow the standard $4D$ internal
evolution and have no gravitational effect on the bulk space.
Stealth branes can nucleate spontaneously in the bulk spacetime.  This
process is described by the standard $4D$ quantum cosmology formalism
with tunneling boundary conditions for the brane world wave function.
The notorious ambiguity in the choice of boundary conditions is fixed
in this case due to the presence of the embedding spacetime.
We also point to some problematic aspects of models admitting stealth
brane solutions.

\end{abstract}

\end{titlepage}

\section {Introduction}

The idea that our Universe could be a (3+1)-dimensional surface
(brane) floating in a higher dimensional embedding (bulk) space dates
back to Regge and Teiltelboim \cite{Regge}. They considered pure
gravity with a conventional looking action 
\be
S^{(4)} _G = \frac{M_{(4)} ^2}{2}\int_{b} d^4 x \sqrt{-g}{R}
\label{eq:grav4}
\ee
where $M_{(4)}$ is the Planck mass and R is the scalar
curvature. However, the integration in (\ref{eq:grav4}) is performed
over a (3+1)-dimensional surface 
\be
y^A = y^A(x)
\ee
in an N-dimensional embedding spacetime and the metric $g_{\mu \nu}$
is the induced metric on the surface, 
\be
g_{\mu \nu} =
G_{AB}\frac{\partial{y^A}}{\partial{x^{\mu}}}\frac{\partial{y^B}} 
{\partial{x^{\nu}}}.   
\ee
Here, $y^{A}  (A=0,1,...,N-1)$ are the coordinates in the embedding
space and the surface is parametrized by the coordinates $x^{\mu}
(\mu = 0,1,2,3)$. The independent dynamical variables of the theory
are not the metric components $g_{\mu \nu}$ but the embedding
functions $y^A (x)$, with the bulk metric $G_{AB}(y)$ being
fixed. Generalizations have later been considered \cite{Davidson} with a matter term added to the action 
\be
S^{(4)} _m = \int_{b} d^4 x \sqrt{-g}L_m \,\, .
\label{eq:mat4}
\ee
The simplest form of the matter term is a 4D cosmological constant,
$L_m = \rho_v = const$, which correspond to the brane tension term from
the bulk point of view.
 
In most of the recent work on brane cosmology, the bulk geometry is
assumed to be dynamical, while the curvature term on the brane is
omitted. The corresponding action includes the bulk curvature term 
\be
S^{(N)} _G = \frac{M_{(N)} ^{N-2}}{2}\int d^N y \sqrt{-G}{\cal R},
\label{eq:gravb}
\ee
the brane matter term (\ref{eq:mat4}), and the bulk matter term
\be
S^{(N)} _m = \int d^N y \sqrt{-G}{\cal L}_m.
\label{eq:matb}
\ee
The latter term is usually taken in the form of a bulk cosmological
constant, which may take different values on the two sides of the
brane. The 4D gravity on the brane is then recovered either by
compactifying the extra dimensions \cite{Arkani} or by introducing a
large negative cosmological constant in the bulk, which causes the
bulk space to warp, confining low-energy gravitons to the brane
\cite{Randall}. 

Quite recently, Dvali, Gabadadze and Porrati (DGP)
\cite{Dvali1} have pointed out that 4D gravity can be recovered even
in an asymptotically Minkowski bulk, provided that one includes the
brane curvature action (\ref{eq:grav4}), 
\be
S= S^{(N)} _{G} + S^{(N)} _m + S^{(4)} _G + S^{(4)} _m.
\label{eq:action}
\ee
Assuming a 5-dimensional bulk and a ${Z_2}$ symmetry of reflections
with respect to the brane, they found that gravity on the brane is
effectively 4D on scales $r<<r_0$, with 
\be
r_0 = \frac{M^2 _{(4)}}{2M^3 _{(5)}},
\label{eq:cross}
\ee
and becomes 5D on larger scales. Analysis of cosmological solutions
with a Robertson-Walker metric on the brane \cite{Deffayet} indicates
the same crossover scale (\ref{eq:cross}) in this class of models. 

The DGP model can be extended in several directions. First, one can
consider 
$N> 5$. One finds that the brane gravity is always 4D for
zero-thickness branes \cite{Dvali2}. However, the crossover to a
higher-dimensional behaviour is recovered when the finite thickness of
the brane is taken into account. 

Another possible extension is to lift the requirement of $Z_2$
symmetry. This symmetry is certainly not mandated by the action
(\ref{eq:action}) and actually cannot be enforced in 5D models where
the brane is coupled to a 5-form field, so that the 5D cosmological
constant is different on the two sides of the brane. Brane world
cosmology without $Z_2$ symmetry has been discussed by a number of
authors
\cite{Binetruy,Ida,Dolezel,Gregory,Davis,Carter1,Anchordoqui,Carter2,Carter3,Holdom1,Holdom2},  
but in most of this work the brane curvature term (\ref{eq:grav4}) has not
been included in the action. Somewhat surprisingly, effective 4D
gravity on the brane could still be obtained for weak sources
\cite{Carter2,Carter3}, but this behaviour does not extend to strong
gravitational fields or to the early Universe cosmology. Some effects
of including the brane curvature term have been discussed in
\cite{Holdom1,Holdom2}. 

In this paper we are going to discuss brane world cosmology in an
infinite bulk with the brane curvature included in the action and
without the assumption of $Z_2$ symmetry. We will show that inclusion
of the curvature term results in some qualitatively new, interesting,
but potentially problematic features of the model. In particular, we
show that the brane produces no gravitational effect in the bulk,
provided that 4D Einstein's equations are satisfied on the brane.  
The brane can then float without disturbing the embedding space, 
and its internal evolution will be identical to that in the
genuine 4D case. 

Closed branes in this model can be spontaneously created in the
bulk.  
The corresponding instanton is a trivial embedding of the usual $S_4$
instanton of  4D quantum cosmology. 
An important difference, however, is that nucleation of brane worlds
occurs in an embedding spacetime. As a result, the brane world wave
function has a clear interpretation, and the issue of boundary
conditions in quantum brane cosmology can be definitively addressed.  

The paper is organized as follows. In the next Section we demonstrate
the existence of branes that satisfy 4D Einstein's equations and do
not disturb the bulk (we refer to them as ``stealth''
branes). Quantum nucleations of spherical branes is discussed in
Section III. In Section IV we consider the dynamics of more general (not
necessary stealth) spherical branes in a 5-dimensional embedding
space. Here we also allow for a nonzero bulk cosmological constant
which can differ on the two sides of the brane.  In Section V we
discuss small-scale quantum fluctuations of the branes and argue that
these fluctuations would be unacceptably large if we lived on a
stealth brane.  Finally, in Section VI
we summarize our conclusions and discuss the potential problems of
models admitting stealth brane solutions.

\section{Stealth branes}

We consider an action of the following form
\beq
S&=&\int_{M} d^{N}Y \sqrt{-G} \left( \frac{1}{2k}{\cal R} + {\cal
L}_{m}\right) \nn \\ 
&+&  \int_{b} d^4 x \sqrt{-g} \left( \frac{1}{2k'}{R} + {L}_m \right) \, \, , 
\label{eq:div} 
\eeq
where $k=M_{(N)} ^{2-N}$, $k'= M_{(4)} ^{-2}$, $M_{(N)}$ and $M_{(4)}$
being the bulk and the brane Planck masses, respectively. The 4D
integration is over the brane $y^A = y^A(x)$, and the induced metric
on the brane can be expressed as 
\be
g_{\mu \nu} = \int_M d^{N}y G_{AB}(y) \P_\mu y^A (x)\P_\nu y^B (x)
\delta^{N} (y^A - y^A(x)) \, \, . 
\label{eq:metrics}
\ee
Variation of the action (\ref{eq:div}) with respect to $G_{AB}$ gives
\beq
\delta S &=& -\frac{1}{2k} \int_M d^{N}y \sqrt{-G} \left(\R ^{AB} -
\frac{1}{2} G^{AB} \R - k{\cal T}^{AB} _{bulk} \right)\delta G_{AB}
\nn \\ 
&{}&-\frac{1}{2k'} \int_b d^{4}x \sqrt{-g} \left( R ^{\mu \nu} -
\frac{1}{2} g^{\mu \nu} \R  - k'{T}^{\mu \nu}\right) \delta g_{\mu
\nu}\, \, , 
\label{eq:var} 
\eeq
where 
\beq
{\cal T}^{AB} _{bulk} &=& \frac{2}{\sqrt{-G}}\frac{\delta{\cal
L}_M}{\delta G_{AB}} \, \, , \nn \\ 
{T}^{\mu \nu} &=& \frac{2}{\sqrt{-g}}\frac{\delta{L}_M}{\delta g_{\mu
\nu}} \,\, , 
\eeq
are the bulk and the brane energy-momentum tensors, respectively, and
the variation $\delta g_{\mu \nu}$ is to be expressed in terms of
$\delta G_{AB}$ using the relation (\ref{eq:metrics}). We thus obtain
the N-dimensional Einstein's equations 
\be
{\cal R}^{AB} - \frac{1}{2}G^{AB}{\cal R} = k( {\cal T}^{AB} _{bulk} +
T^{AB} _{brane}) 
\label{eq:einstein} \,\, ,
\ee 
where
\be
T_{brane} ^{AB}= \frac{1}{\sqrt{-G}} \int_b d^4 x \sqrt{-g} \left(
T^{\mu \nu} - \frac{1}{k'}\left(R^{\mu \nu} - \frac{1}{2}g^{\mu \nu}R
\right)\right)\P_\mu y^A(x) \P_\nu y^B(x)\delta^{N}(Y^A -Y^A (x)) \,
\, . 
\label{eq:energy}
\ee

From equations (\ref{eq:einstein}), (\ref{eq:energy}) we see
immediately that if 4D Einstein's equations are satisfied on the
brane, 
\be
{\tilde T}^{\mu \nu} \equiv  T^{\mu \nu} - \frac{1}{k'}\left(R^{\mu
\nu} - \frac{1}{2}g^{\mu \nu}R \right)=0 \,\, , 
\label{eq:enten}
\ee
then the brane has no gravitational effect on the N-dimensional
bulk.  The brane can then be treated as evolving in a fixed background
geometry. We shall refer to such branes as ``stealth'' branes.  

The brane equations of motion can be obtained by varying
the action with respect to the embedding functions $y^A(x)$. These
equations can be expressed as \cite{Carter4} 
\be
{\tilde T}^{\mu \nu} K_{\mu \nu} ^i = 0
\label{eq:acc}
\ee
where ${\tilde T}^{\mu \nu}$ is from equation (\ref{eq:enten}),
\be
K^i _{\mu \nu} = - n_A ^i D_\mu  e^A _\nu
\ee
is the extrinsic curvature, $n^i$ are the (N-4) unit normal vectors to
the worldsheet with tangent vectors $e_\mu^A = y^A _{,\mu}$, and
$D_\mu = e^A _\mu \nabla_A$, with $\nabla_A$ being the covariant derivative
in the metric $G_{AB}$. 

Suppose now that we have a solution of 4D Einstein's equations
(\ref{eq:enten}) which can be embedded into a N-dimensional Minkowski
space with some functions $y^A (x)$. It is then clear that such an
embedding gives a solution both of the N-dimensional Einstein's
equations (\ref{eq:einstein}) (N-dimensional Minkowski space with
${\cal T}^{AB}=0$) and of the brane equations of motions
(\ref{eq:acc}). Thus any 4D Universe which is embeddable into
Minkowski bulk represents a possible internal evolution of a brane
world.\footnote{A special case of this general result was noted by Dick
\cite{Dick}, who pointed out that a spatially flat 3-brane in a 5D
bulk without
$Z_2$ symmetry may follow the standard Friedman evolution.}

For $N\geq 10$, any 4D Universe can be embedded at least locally
\cite{Janet}, and for $N\geq 91$. a global embedding is also possible
\cite{Clarke}. Any solution of 4D Einstein's equations can then be
realized as a brane world evolution. 

In most of the recent work on brane worlds it is assumed that the bulk
is 5-dimensional, in which case there are significant restrictions on
the possible embeddings. For example, the Schwarzschild solution can
only be embedded for $N\geq 6$ \cite{Kasner}. However, closed and flat
Friedman-Robertson-Walker cosmologies are all embeddable in 5D
Minkowski space. 

We assumed so far that stealth branes evolve in a Minkowski
bulk. However, it is clear from equations
(\ref{eq:einstein})-(\ref{eq:acc}) that one can use practically any
fixed background. In particular, we shall later consider stealth
branes in de Sitter and anti-de Sitter spaces.

\section{Nucleation of stealth branes}

From the point of view of the bulk, stealth branes have vanishing
 energy and momentum, and their nucleation is not forbidden by any
 conservation laws. One can imagine small, nearly spherical branes
 filled with a high-energy false vacuum nucleating in Minkowski
 space. The branes then go through a period of inflation, thermalize,
 and evolve (locally) along the lines of the standard hot cosmological
 model. This picture is identical to that usually adopted in quantum
 cosmology \cite{Vilenkin1,Hawking,Linde,Vilenkin2,Rubakov}, except for the
 presence of the embedding space\footnote{Note that our discussion
 here differs from \cite{Garriga1}  who considered nucleation of 5D
 Universes consisting of AdS regions joined along a brane.}. 
 One can expect that the nucleation rate is described by the same
 4-sphere instanton, which is now embedded in a Minkowski bulk. The
 bulk does not contribute to the instanton action, so the action is
 also the same, 
\beq
S^{(E)} &=& \int d^4 x \sqrt{g}(-\frac{1}{2k'}R + \rho_V) \nn \\
&=& - 24\pi^2 M^4 _{(4)}/\rho_V ,
\label{eq:ins1}
\eeq
where $\rho_V$ is the 4D false vacuum energy density. 

The role of the vacuum energy in inflationary models is usually played
by a scalar field potential $V(\varphi)$. This potential is assumed to
be very flat, at least in some range of $\varphi$, so that $\varphi$
evolves very slowly and $V(\varphi)$ acts as a vacuum energy. Then,
the instanton action 
\be
S^{(E)} (\varphi) = -24\pi^2M^4 _{(4)}/V(\varphi)
\label{eq:insin}
\ee
should determine the relative probability of nucleation for branes
with different values of $\varphi$. 

The situation is complicated, however, by the ongoing debate
\cite{Vilenkin10} about how the instanton action (\ref{eq:insin})
appears in the nucleation rate ${\cal P}$. Different results are
obtained depending on one's choice of the boundary conditions for the
wave function of the Universe. The Hartle-Hawking \cite{Hawking} wave
function gives 
\be
{\cal P}_{HH} \propto \exp(-S^{(E)}) \,\, ,
\label{eq:hh}
\ee 
while the tunneling \cite{Vilenkin2,Rubakov} and Linde \cite{Linde}
wave functions give 
\be
{\cal P}_{T} \propto \exp(-|S^{(E)}|) \,\, .
\label{eq:li}
\ee 
Since the instanton action (\ref{eq:insin}) is negative, the difference
between (\ref{eq:hh}) and (\ref{eq:li}) is quite dramatic. This issue
has not found a clear resolution in 4D quantum cosmology, mainly due
to the conceptual problems with the interpretation of the wave
function of the Universe. We believe the brane world picture can shed
some new light on this problem, since the wave function of a brane
nucleating in a Minkowski bulk has a clear physical interpretation. 

Quantum cosmology of spherical branes in a 5D Minkowski space has been
studied by Davidson et. al. \cite{Karasik1,Karasik2} in the framework of
Regge-Teitelboim theory. For stealth branes in our model, the bulk
gravitational field is absent and the model of
\cite{Karasik1,Karasik2} is an appropriate minisuperspace
approximation. The 5D metric can be written as  
\be
ds^2 _5 = \sigma ^2 (-d\tau ^2 +da^2 +a^2d\Omega ^2 _3)\,\, ,
\ee  
where $d\Omega ^2 _3$ is the metric on a unit 3-sphere and $\sigma ^2$
is a normalization factor. The evolution of the brane is described by
specifying the functions $a(t)$ and $\tau(t)$, where $t$ is a time
parameter on the brane. We shall choose $t$ to be the proper time,
which introduces the gauge condition 
\be
{\dot{\tau}}^2 - {\dot{a}}^2 =1,
\ee
where dots stand for derivatives with respect to $t$. 

With a suitable choice of the normalization,
$\sigma^2=(12\pi^2M_{(4)}^2)^{-1}$, the conserved energy of the brane
is given by 
\be
{\cal E} = \frac{1}{2}( {\dot{a}}^2 + 1 -H^2 a^2)a\sqrt{1 + {\dot{a}}^2},
\ee
where
\be
H^2 = \rho_V(6\pi M^2 _{(4)})^{-2}.
\ee
In the case of a stealth brane ${\cal E} =0$,
\be
{\dot a}^2 +1 - H^2 a^2 =0,
\ee
and one recovers the usual 4D Wheeler-DeWitt equation
\be
[ p^2 _a + a^2(1-H^2 a^2)]\psi =0
\label{eq:WDW}
\ee
where
\be
p_a = -a\dot{a}
\label{eq:mom}
\ee
is the momentum conjugate to $a$. (We disregard the factor ordering
ambiguities in this discussion). In the ``coordinate'' representation,
the wave function is $\psi = \psi(a)$ and $p_a = -i
\frac{\partial}{\partial a}$.  

Davidson et. al. were not specifically concerned with the problem of
brane nucleation, so they considered branes of arbitrary energy ${\cal
E}$ and a variety of possible boundary conditions at $a \rightarrow
\infty$. They also required that the wave function should vanish at
$a=0$, which in our view is not justified. 

Let us now address the problem of boundary conditions for the
nucleation of stealth branes $({\cal E}=0)$. The Wheeler-DeWitt
equation has the form of a one-dimensional Schroedinger equation for a
``particle'' described by a coordinate $a(t)$, having zero energy, and
moving in a potential 
\be
U(a) = a^2 (1-H^2a^2)
\ee
The classically allowed region is $a \geq H^{-1}$, and the WKB
solutions of equation (\ref{eq:WDW}) in this region are 
\be
\psi_{\pm} (a) = [p(a)]^{-1/2} \exp[ \pm i \int^a _{H^{-1}}p(a')da'
\mp \frac{i\pi}{4}], 
\label{eq:wkb}
\ee
where $p(a) = [-U(a)]^{1/2}$. The three widely discussed choices of
boundary conditions correspond to 
\be
\psi_{T} (a > H^{-1}) = \psi_{-} (a)
\label{eq:wtv}
\ee
for the tunneling wave function, 
\be
\psi_{HH} (a > H^{-1}) = \psi_{+}(a) - \psi_{-} (a)
\label{eq:wthh}
\ee
for the Hartle-Hawking wave function, and
\be
\psi_{L} (a > H^{-1}) = \psi_{+}(a) + \psi_{-} (a)
\label{eq:wtl}
\ee
for the Linde wave function.

For the two semiclassical wave functions (\ref{eq:wkb}) we have
\be
{\hat p}_a \psi_{\pm}(a) \approx \pm p(a) \psi_{\pm}(a) \,\, .
\ee
Combined with equation (\ref{eq:mom}), this indicates that $\psi_{-}
(a)$ and $\psi_{+}(a)$ describe an expanding and a contracting
Universe, respectively. Hence, the tunneling wave function includes
only the expanding component, while the Hartle-Hawking and Linde wave
functions include both components with equal weight and appear to
describe contracting and re-expanding Universes. 

This interpretation has been questioned by Rubakov \cite{Rubakov2} who
notes that the time coordinate $t$ is just an arbitrary label, so
changing expansion to contraction does not do anything, as long as the
directions of all other physical processes are also reversed. In our
simple minisuperspace model $a(t)$ is the only variable, and thus
$\psi_{+} (a)$ and $\psi_{-} (a)$ may both correspond to an expanding
Universe. 

Rubakov's objection does not, however, apply to the case of nucleating 
branes. In this case we have
\be
{\dot a} = \frac{da}{d\tau}\sqrt{1 + {\dot a}^2},
\label{eq:vel}
\ee
and thus the brane expanding in terms of its proper time $t$ is also
expanding in terms of the Minkowski time $\tau$. We could of course
reverse the internal time coordinate, which would introduce a minus
sign on the right-hand side of equation (\ref{eq:vel}). But still, one
of the wave functions (\ref{eq:wkb}) would correspond to an expanding
and the other to a contracting brane, in terms of the Minkowski time
$\tau$. In the brane nucleation process, the newly created branes
expand and no contracting branes are present. Thus the tunneling wave
function (\ref{eq:wtv}) (or its complex conjugate) is the only correct
choice in this case. 

Stealth brane nucleation can also occur in curved spacetime. Important 
examples here are de Sitter and anti-de Sitter spaces.
The instanton describing stealth brane nucleation in de Sitter space
is the same 4-sphere instanton as in Minkowski space, except now it is
embedded into an $N$-sphere (Euclideanized $N$-dimensional de Sitter
space).  The difference between the instanton action and the $N$-sphere
action without a brane is still given by Eq. (\ref{eq:ins1}), so the
rate of brane nucleation is also unchanged.

Turning now to anti-de Sitter space, we consider the case of a single
extra dimension for simplicity. The Euclideanized anti-de Sitter space
metric can be written as (see, e.g., \cite{Garriga1})
\be
ds^2 _{E} = dr^2 + l^2 \sinh ^2 (r/l)[ d\chi ^2 + \sin ^2 \chi d\Omega^2 _3 ],
\ee
where $l = (-6/\Lambda)^{1/2}$ and $\Lambda < 0$ is the 5D cosmological
constant. The brane worldsheet is a sphere $r=r_0$, where $r_0$ is
determined from 
\be
H^{-1} = l\sinh (r_0/l).
\ee
The Lorentzian evolution is obtained by the analytic continuation 
$\chi \rightarrow iHt + \pi/2$,
\be
ds^2 = dr^2 + (lH)^2 \sinh ^2 (r/l)[- dt^2 + H^{-2}\cosh ^2 (Ht) 
d\Omega^2 _3 ],
\ee
with the brane worldsheet still given by $r=r_0$.

\section{Spherical branes in 5D}

We now turn to a more general situation, when 4D Einstein's equations
on the brane are not necessarily satisfied. Here, we shall specialize
to the case of spherical branes and of a 5D bulk. We shall also assume
a non-zero bulk cosmological constant and allow for the possibility
that it can take different values on the two sides of the brane. 

Following the same steps as in \cite{Ida}, the equations of motion of
the brane can be expressed as     

\beq
[K]g_{\mu \nu} - [K_{\mu \nu}] &=& k{\tilde T}_{\mu \nu}, 
\label{eq:jc} \\
{\tilde T}^{\mu \nu} <K_{\mu \nu}> &=& [{\cal T}_{nn}],
\label{eq:force}\\
\nabla_\nu (T_\mu ^\nu) &=& - [{\cal T}_{\mu n}] 
\label{eq:conservation}
\eeq 
Here, $K_{\mu \nu}$ is the extrinsic curvature of the brane, ${\cal
T}_{nn}= ({\cal T}_{bulk})_{AB}n^A n^B$, ${\cal T}{}_{\mu n}= ({\cal
T}_{bulk})_{AB}e^A _\mu n^B$, the square and angular brackets stand,
respectively, for the difference and the average of the corresponding
quantity on the two sides of the brane, e.g., $[K_{\mu \nu}]= K_{\mu
\nu} ^{+} - K_{\mu \nu} ^{-}$, $<K_{\mu \nu}>= \frac{1}{2}(K_{\mu \nu}
^{+} + K_{\mu \nu} ^{-})$, where ``$+$'' and ``$-$'' correspond,
respectively, to the brane exterior and interior. 
\footnote{The exterior and interior are identical for $Z_2$ symmetric
branes, in which case we either have two spherical 4D regions joined
at the brane, or two possibly infinite regions connected by a
wormhole. Here we are interested in the case of trivial topology, with
spherical interior region and an infinite region outside.}   

We shall assume that the bulk energy-momentum tensor has the vacuum form,
\be
{\cal T}^{\pm} _{AB} = - k^{-1}\Lambda^{\pm} g_{AB} \,\, .
\ee
Then, using the generalized Birkhoff theorem \cite{Gregory}, the 5D
metric can be expressed as 
\be
ds^2 _5 = -A_{\pm}d\tau^2 + A_{\pm} ^{-1}da^2 + a^2d\Omega^2 _3 
\label{eq:Birkhoff}
\ee
with
\be
A_{\pm} = 1 - \frac{\Lambda^{\pm}}{6}a^2 - \frac{2{\cal M}^{\pm}}{M^3
_{(5)}a^2} \,\, . 
\label{eq:ms}
\ee
In the proper time gauge, the metric on the brane is
\be
ds_4 ^2 = -dt^2 + a^2 (t)d\Omega^2 _3 \,\, .
\label{eq:mc}
\ee
The brane embedding is described by specifying the functions $ a(t)$,
$\tau (t)$. 

The components of the extrinsic brane curvature of the worldsheet are
\beq
&{}&K^{\pm}_{\tau \tau} = -\frac{\left( \ddot{a} + \frac{1}{2}
\frac{\partial A_{\pm}}{\partial a} \right)}{\left( \dot{a} ^2 +
A_{\pm} \right)^{1/2}} \\  
&{}& K^{\pm \chi} _{\chi}=K^{\pm \theta} _{\theta} = K^{\pm \phi}
_{\phi}=\frac{( {\dot{a}}^2 + A_{\pm})^{1/2}}{a} \,\, . 
\eeq
Substituting this in the junction condition equations (\ref{eq:jc})
and assuming that the brane energy-momentum tensor is of the form 
\be
T^{\nu} _{\mu} = {\mbox{diag}}(\rho, -P, -P, -P) \,\, ,
\ee
one obtains \cite{Holdom1}
\be
\left( \dot{a}^2 + A_{-} \right)^{1/2} - \left( \dot{a}^2 + A_+
\right)^{1/2}= \frac{ka}{3}\left( \rho - \frac{3(\dot{a} ^2 +
1)}{k'a^2} \right).   
\label{eq:jcs}
\ee
Equation (\ref{eq:conservation}) expresses the energy-momentum
conservation on the brane, 
\be
\dot{\rho} + 3\frac{\dot{a}}{a}(\rho + P) = 0
\label{eq:cons}
\ee
while equation (\ref{eq:force}) gives a relation which is obtainable
from (\ref{eq:jcs}) and (\ref{eq:cons}). 

A stealth brane corresponds to ${\cal M}^{\pm}=0$ and $\Lambda^+ =
\Lambda^{-} \equiv \Lambda$. As expected, in this case
Eqs. (\ref{eq:jcs}), (\ref{eq:cons}) reduce to the standard FRW
evolution equations. Even if the bulk cosmological constant is
non-zero, it has no effect on the evolution of the brane. 

It should be noted that, in the case of a positive $\Lambda$, the
static coordinates (\ref{eq:Birkhoff}) and (\ref{eq:ms}) cannot be
used beyond the de Sitter horizon, $a< {\cal H}^{-1}$, where ${\cal
H}= (\Lambda/6)^{1/2}$. 

\noindent However, since no singularities are developed at $a={\cal
H}^{-1}$, one can expect that the solution for $a(t)$ will be
analytically continued, so that the standard evolution will continue
at $a>{\cal H}^{-1}$. 

We now briefly consider some examples of deviations from stealth brane
evolution. 

Suppose first that the bulk cosmological constant vanishes,
$\Lambda^{\pm}=0$, but the brane has a nonzero mass, ${\cal
M}^+\equiv {\cal M}\not=0$, with ${\cal M}^-=0$.  
For sufficiently large brane radii,
$a^2\gg{\cal M}/M_{(5)}^3$, the left-hand side of (\ref{eq:jcs}) can
be expanded to yield
\be
{{{\dot a}^2+1}\over{a^2}}={1\over{3M_{(4)}^2}}\left(\rho-{{\cal
M}\over{a^3\sqrt{{\dot a}^2+1}}}\right).
\label{dav}
\ee
The second term in the brackets represents a correction to the stealth
evolution due to a non-zero brane mass.  (Note that the mass ${\cal
M}$ can be both positive or negative.)

The same equation (\ref{dav}) is obtained in the Regge-Teitelboim
limit of vanishing bulk gravity, $M_{(5)}\to\infty$.  The brane
dynamics in this regime has been investigated by Davidson
\cite{Davidson}. 

Suppose now that ${\cal M}^{\pm}=0$, while $\Lambda^+$ and $\Lambda^-$
are both non-zero, and that the brane has a vacuum equation of state,
$\rho= {\rm const}$.  Then it is easily understood that
Eq. (\ref{eq:jcs}) has a de Sitter solution,
\be
a(t)=H^{-1}\cosh(Ht),
\ee
where $H$ can be found from the equation
\be
\left( H^2-{\Lambda^-\over{6}}\right)^{1/2}-
\left( H^2-{\Lambda^+\over{6}}\right)^{1/2}={1\over{M_{(5)}^3}}
\left({\rho\over{3}} -M_{(4)}^2H^2\right).
\label{HLpm}
\ee
The instanton describing nucleation of such branes consists of de
Sitter 5-spheres of radii $H_+ ^{-1}  = (6/\Lambda^+)^{1/2}$ and $ H_{-} ^{-1}  = (6/\Lambda^-)^{1/2}$ 
joint at a 4-sphere of radius $H^{-1}$ with $H$ from Eq. (\ref{HLpm}).

The corresponding Euclidean action can be expressed as
\beq
S^{(E)} &=& -\frac{8\pi^2}{H^2}M_{(4)}^2  - 4\pi^2\frac{M_{(5)}^3}{H^2}\left( \frac{(H^2 - H_+ ^2)^{1/2}}{H_+ ^2}- \frac{(H^2 - H_{-} ^2)^{1/2}}{H_{-} ^2} \right) \nn \\  
&-& 4\pi^2 M_{(5)}^3 \left(\frac{\phi_{-}}{H_{-}^3} - \frac{\phi_{+}}{H_{+}^3} \right)
\eeq
where $\sin \phi_+ = \frac{H_+}{H}$ and $\sin \phi_{-} = \frac{H_{-}}{H}$.
This cumbersome expression can be simplified in the limit when
$H_{\pm} \ll H$,
\be
S^{(E)} \approx - \frac{24 \pi^2M_{(4)} ^4}{\rho} - \frac{16\pi ^2}{5H^5} M_{(5)}^3 (H_+ ^2 - H_{-} ^2)
\ee
with $H^2 \approx \rho/3M_{(4)}^2$.

\section{Do we live on a stealth brane?}

Our discussion in sections II and III naturally leads to the question:
Is it possible that we live on a stealth brane? From the classical
point of view, this is quite possible, since the classical interior
evolution of stealth branes is identical to that of genuine 4D
universes. We shall now argue, however, that quantum-mechanically this
picture leads to physically unacceptable consequences. 
 
If we live on a stealth brane, then the tension of the brane is equal to the 4D cosmological constant. This is observationally constrained by
\be
\rho_v \alt (10^{-3} eV)^4 .
\label{eq:bound}
\ee
We are going to argue that with such extremely small tension, the quantum fluctuations of the brane on small length scales would be unacceptable large. Moreover, the scalar particles corresponding to these fluctuations would be copiously produced in high-energy particle collisions in accelerators.

Small fluctuations on a 3-brane in an N-dimensional bulk spacetime can be described by a set of $(N-4)$ scalar fields with the Lagrangian \cite{Vilenkin11}
\be
-\frac{1}{2} \rho_v ( Y^a _{,\mu} Y^{a,\mu} + RY^a Y^a ).
\label{YY}
\ee
Here, $Y^a (x)$ have the meaning of brane displacements in $(N-4)$
orthogonal directions and we assume summation over $a=1,..,N-4$. For a
nearly flat brane, the curvature term in (\ref{YY}) 
is negligible, and we have a set of massless scalar fields. 

The characteristic amplitude of zero-point brane fluctuations on a
length scale $L$ can be estimated by requiring that the corresponding
action in a spacetime volume $\sim L^4$ is $\sim 1$, 
\be
\rho_v (Y^a/L)^2 L^4 \sim 1.
\ee
This gives 
\be
Y^a \sim \rho^{-1/2} _v L^{-1}
\ee
The brane can be treated as a well-localized classical object as long
as $Y^a \ll L$, that is, on scales $L \gg \rho^{-1/4} _v$. On smaller
scales, the brane inhabitants would find that the classical picture of
spacetime no longer applies. With $\rho_v$ satisfying the bound
(\ref{eq:bound}), deviations from the classical picture would be seen
on scales as large as 1 mm. 

A related problem is that, for small values of $\rho_v$, quanta of the
fields $Y^a$ can be copiously created in particle
collisions.\footnote{We are grateful to Gia Dvali for pointing this
out to us.} Consider
for example the coupling of $Y^a$ to the electromagnetic field $F_{\mu
\nu}$. The canonically normalized fields corresponding to $Y^a$ are
$\psi^a = \rho^{1/2} _v Y^a$. In terms of these fields, the 4D metric
is 
\be
g_{\mu \nu} = \eta_{\mu \nu} + \rho^{-1} _v  \psi^a _{,\mu}\psi^a _{,\nu},
\ee
and the interaction Lagrangian is
\be
\rho^{-1} _v \left(\psi^a _{,\mu}\psi^a _{,\nu} - \frac{1}{4}\eta_{\mu
\nu}\psi^a _{,\sigma}\psi^{a,\sigma}\right) F^{\mu \rho} F^\nu _{\rho}
. 
\ee
The effective coupling constant at energy $\epsilon$ is $\lambda_{eff}
\sim \epsilon^4 /\rho_v$, so the theory becomes strongly interacting at
energies $\epsilon \agt \rho^{1/4} _v$. This is in conflict with
observations, unless 
\be
\rho_v \agt (1\mbox{TeV})^4 .
\label{eq:bound1}
\ee
The conclusion is that in brane world models without $Z_2$ symmetry the brane tension must satisfy the bound (\ref{eq:bound1}).
These models include the Randall-Sundrum-type 5D model with a positive
brane tension $\rho_v$, a negative bulk cosmological constant
$\Lambda$, and a brane curvature term in the action. 
The values of $\rho_v$ and $\Lambda$ are tuned to allow (nearly) flat
brane solutions. The original Randall-Sundrum model assumed $Z_2$
symmetry, but it can be lifted, as long as (\ref{eq:bound1}) is
satisfied. Then, we may be living on a flat brane separating two
anti-de Sitter regions, but stealth branes will nucleate and expand in
the anti-de Sitter bulk. Their internal geometry will be de Sitter,
with the vacuum energy given by the brane tension.  The expanding
stealth branes will occasionally run into our brane, and one can derive
constraints on the model parameters by requiring that such brane
collisions should be sufficiently rare. This has been done in
$\cite{Perkins}$ for a somewhat different model. 

Another possibility is a model with a flat 4D brane world in two extra dimensions. The brane evolution equation (\ref{eq:acc}) is trivially satisfied for $K^i _{\mu \nu} = 0$, and the bulk Einstein's equations (\ref{eq:einstein}) are solved by a conical metric in the extra dimensions. The bulk metric is then locally flat, and stealth brane nucleation will proceed as in flat spacetime. To avoid an excessive rate of collisions of our brane with stealth branes, one can introduce a small cosmological constant in the bulk, and brane collisions will be rare, provided that the brane nucleation rate is sufficiently low. 

\section{Conclusions and discussion}

We discussed the classical and quantum cosmology of 3-branes with an
Einstein curvature term added to the brane action. If $Z_2$ symmetry
between the two sides of the brane is not imposed, the model admits
the existence of stealth branes which follow the standard 4D internal
evolution and have no gravitational effect on the bulk space. Stealth
branes have vanishing energy in the bulk space, ${\cal E} =0$, and can
therefore nucleate spontaneously. This process is described by the
standard 4D quantum cosmology formalism with tunneling boundary
condition for the brane world wave function. The notorious ambiguity
in the choice of the boundary conditions is fixed in this case due to
the presence of the embedding spacetime. 

Apart from stealth branes, the model also allows the existence of
branes with negative bulk energy. So one could have spontaneous
creation of brane pairs, one with energy ${\cal E}_1>0$ and the other
with ${\cal E}_2 = -{\cal E}_1$. Alternatively, a positive energy
brane can increase its energy by chopping off a negative energy
brane. 

We have argued that we are not likely to live on a stealth brane, since otherwise the tension of our brane would be extremely small, $\rho_v \alt (10^{-3} eV)^4$. As a result, the brane would be subject to large quantum fluctuations, and the classical picture of spacetime would break down for the brane inhabitants on unacceptably large length scales. Quanta of brane fluctuations would also be copiously produced in high-energy collisions in particle accelerators. But even though we do not live on a stealth brane, such branes can still nucleate and expand in the bulk spacetime.  

The existence of stealth branes indicates that the
bulk space is unstable, and there is a danger that it may
fill up with nucleated branes, so that the brane inhabitants will be
constantly bombarded by other branes hitting them from extra
dimensions. One way around this problem is to enforce the $Z_2$
symmetry, thus disallowing stealth branes.  
Alternatively, one can try to impose constraints on the parameters of
the model, e.g., an upper bound on stealth brane nucleation rate, to
reduce the rate of brane collisions to an acceptable value. One could
also consider an expanding de Sitter bulk. The branes will then be
driven apart by the exponential expansion of the bulk, and brane
collisions will be rare, provided that the brane nucleation rate is
sufficiently low.  

We finally point to a
disturbing property of stealth brane solutions which is manifested
in cases when the brane spacetime allows a continuum of embeddings.
Consider for example a flat brane 
\be
ds_4^2=-dt^2+dx^2+dy^2+dz^2
\ee
embedded into a flat 5D Minkowski space
\be
ds_5^2=-d\tau^2+dx^2+dy^2+dz^2+dw^2.
\ee
A possible embedding is given by any functions $\tau(t)$ and $w(t)$
satisfying
\be
{\dot\tau}^2-{\dot w}^2=1.
\label{tauw}
\ee  
In particular, we could have $\tau=t$, $w=0$ at $t<0$ and an arbitrary
function $w(t)$ at $t>0$, with $\tau(t)$ determined from (\ref{tauw}).
For example, the brane could suddenly start moving with an
acceleration, $w(t)=at^2/2$.
This shows that the evolution of the brane is degenerate, in the sense
that it is not uniquely predicted
by the initial data.\footnote{We are grateful to Jaume Garriga for
pointing this out to us.}   

One could try to resolve this
problem by 
considering small perturbations about stealth brane solutions.  
The degeneracy of brane evolution may be removed if only
perturbatively stable stealth solutions are allowed (that is,
solutions which admit a full spectrum of infinitesimal perturbations).
Consider, for example, a brane which suddenly starts to accelerate in
a flat $5D$ embedding space.  For perturbations violating Einstein's
equations on the brane, the $5D$ energy of the brane is generally
non-zero, and its accelerated motion in flat space is inconsistent
with energy and momentum conservation.  An accelerating brane solution
can be obtained by introducing an appropriate gravitational field in
the bulk.  However, an infinitesimal perturbation of the bulk
Minkowski geometry can result only in an infinitesimal acceleration
of the brane.  Thus, a brane which suddenly starts moving with a
finite acceleration is not perturbatively stable.

A detailed analysis of this issue is beyond the scope of the present paper.

\section{Acknowledgements}

We are grateful to Gia Dvali and Jaume Garriga for many enlightening
discussions and to Takahiro Tanaka for pointing out an error in the
original version of the paper.
The work of R.C. was supported by CONACYT (postdoctoral fellowship number 148623), SNI (M\'exico) and COTEPABE-IPN, 
and the work of
A.V. was supported in part by the National Science Foundation.

\end{document}